\documentclass[aps,prl,reprint,groupedaddress,showpacs]{revtex4-1}
\usepackage{graphicx}
\usepackage{amssymb}
\usepackage[normalem]{ulem}
\usepackage{xcolor}

\graphicspath{{./IMAGES/}}
\begin{document}

\title{Percolation transition by random vertex splitting of diamond networks?}
\author{Susan Nachtrab}
\author{Matthias J.F.~Hoffmann}
\author{Sebastian C.~Kapfer}
\author{Gerd~E.~Schr\"oder-Turk}
\email{Gerd.Schroeder-Turk@fau.de}
\author{Klaus~Mecke}
\email{Klaus.Mecke@physik.uni-erlangen.de}
\affiliation{Institut f\"ur Theoretische Physik, Friedrich-Alexander Universit\"at Erlangen-N\"urnberg, Staudtstr.~7, D-91058 Erlangen, Germany}

\date{\today}

\begin{abstract}
We propose a statistical model defined on the three-dimensional diamond network where the splitting of randomly selected nodes leads to a spatially disordered network, with decreasing degree of connectivity. The terminal state, that is reached when all nodes have been split, is a dense configuration of self-avoiding walks on the diamond network. Starting from the crystallographic diamond network, each of the four-coordinated nodes is replaced with probability $p$ by a pair of two edges, each connecting a pair of the adjacent vertices. For all values $0\le p \le 1$ the network percolates, yet the fraction $f_p$ of the system that belongs to a percolating cluster drops sharply at $p_c=1$ to a finite value $f_p^{c}$. This transition has the signature of a phase transition with scaling exponents for $p\rightarrow p_c$ that are different from the critical exponents of the second-order phase transition of standard percolation models. As is the case for percolation transitions, this transition significantly affects the mechanical properties of linear-elastic realisations, obtained by replacing edges with solid circular struts to give an effective density $\phi$. Finite element methods demonstrate that, as a low-density cellular structure, the bulk modulus $K$ shows a cross-over from a compression-dominated behaviour, $K(\phi)\propto \phi^\kappa$ with $\kappa\approx 1$, at $p=0$ to a bending-dominated behaviour with $\kappa\approx 2$ at $p=1$. 
\end{abstract}

\pacs{64.60.ah;  %
      64.60.aq;  %
      62.20.F-;  %
}

\maketitle 

{\em Percolation} is a fundamental model of statistical physics and probability theory \cite{StaufferAharony:1994}, with a wealth of scientific and engineering applications \cite{sahimi:1994}. The fundamental question of percolation theory is the existence of connected components whose size is of the order of the system size ({\em percolating clusters}), in disordered structures that result from randomly inserting or removing local structural elements. It owes its generality, and hence importance, partially to the strong universality of the percolation transition. In the majority of lattice and continuum models, the transition from non-percolating to percolating structures is a continuous second-order phase transition in the insertion (or deletion) probability $p$, characterised by the same critical exponents that are independent of lattice type, symmetry, coordination, particle shape, etc \cite{StaufferAharony:1994}. Exceptions are non-equilibrium directed percolation models \cite{Hinrichsen:2000,Odor:2004} and negative-weight percolation \cite{MelchertHartmann:2008}, both with different critical exponents, and explosive percolation where a bias for the formation of small clusters leads to a first order transition \cite{AchlioptasDSouzaSpence:2009,*Ziff:2009} or at least to unusual finite size scaling \cite{GrassbergerChristensenBizhaniSonPaczuski:2011}. 

\begin{figure}[t]
\begin{center}
  \hspace*{0.0\columnwidth}
  \includegraphics[width=0.44\columnwidth]{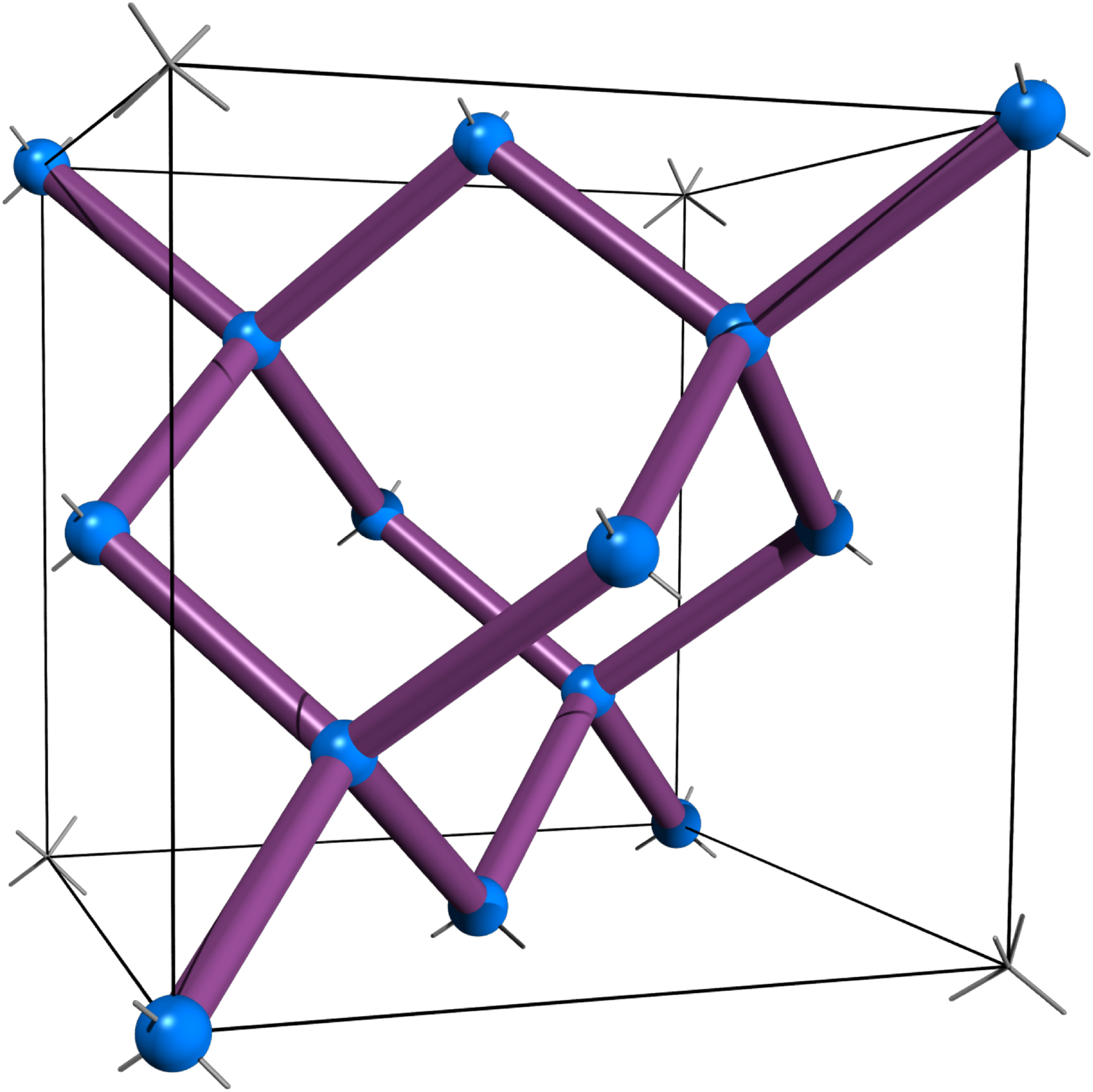}
  \nolinebreak \hfill
  \includegraphics[width=0.44\columnwidth]{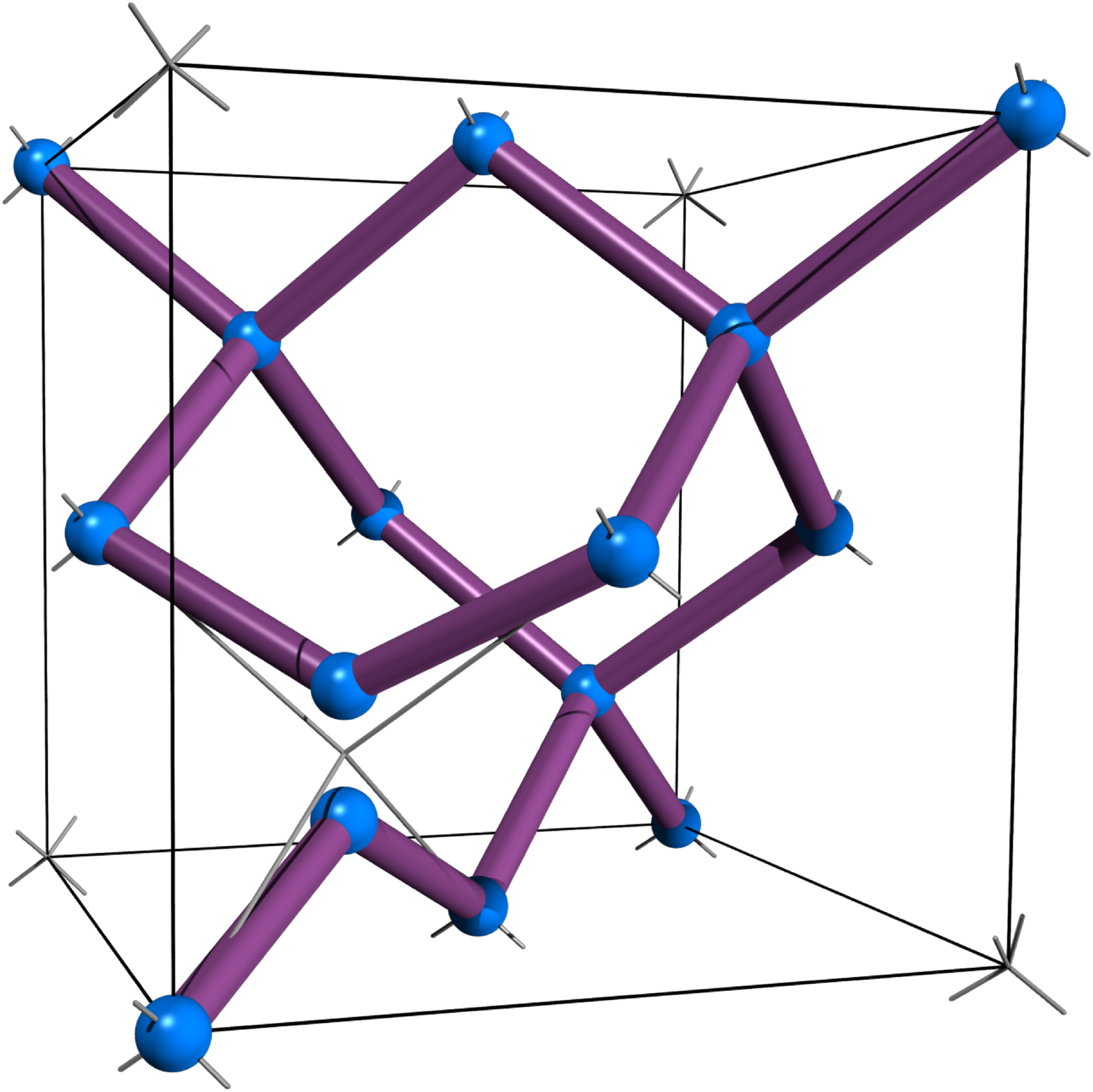}
  \hspace*{0.0\columnwidth}
  \\[0.2cm]
  \includegraphics[width=0.3\columnwidth]{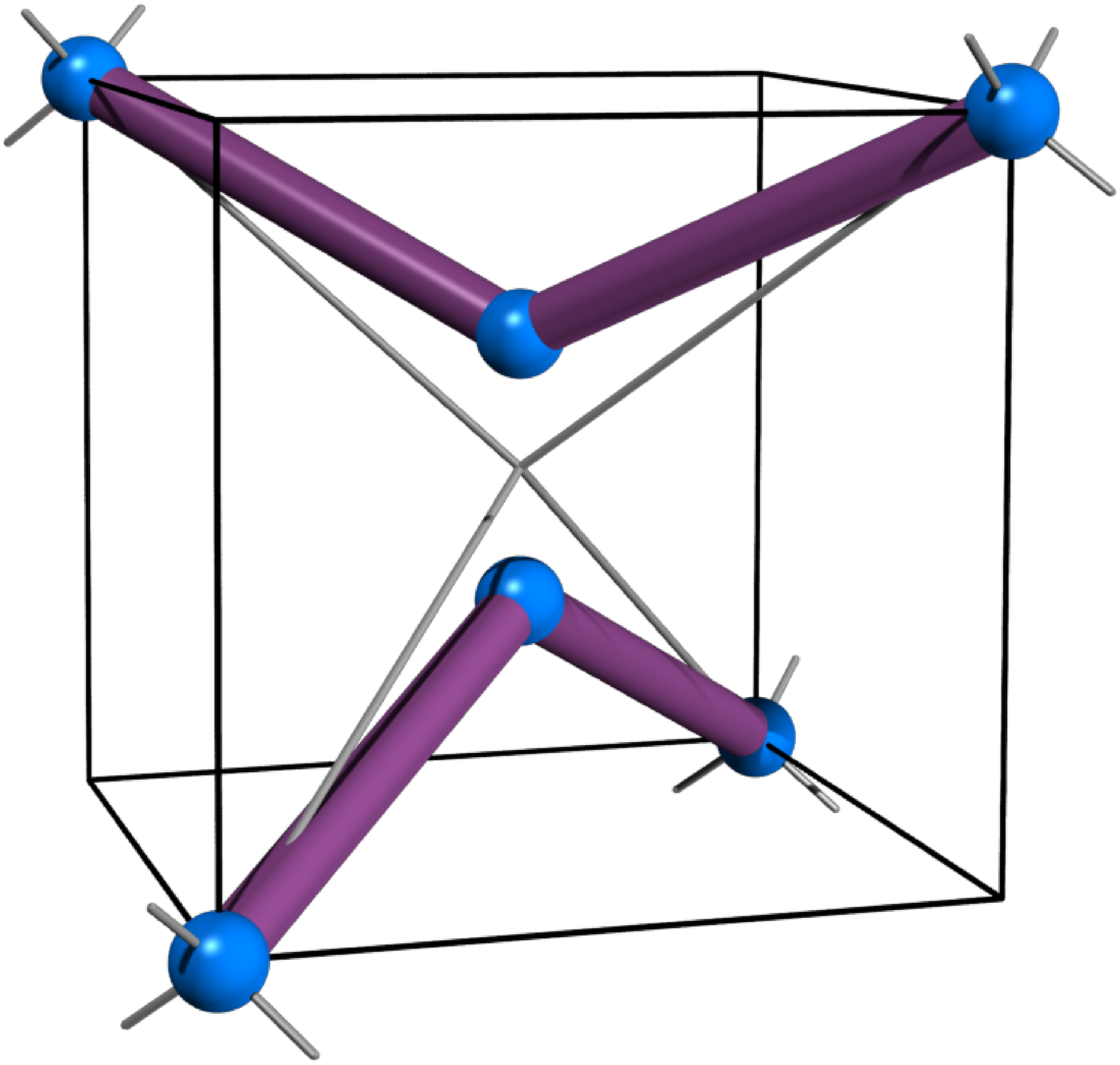}\nolinebreak\hspace*{0.01\columnwidth}
  \includegraphics[width=0.3\columnwidth]{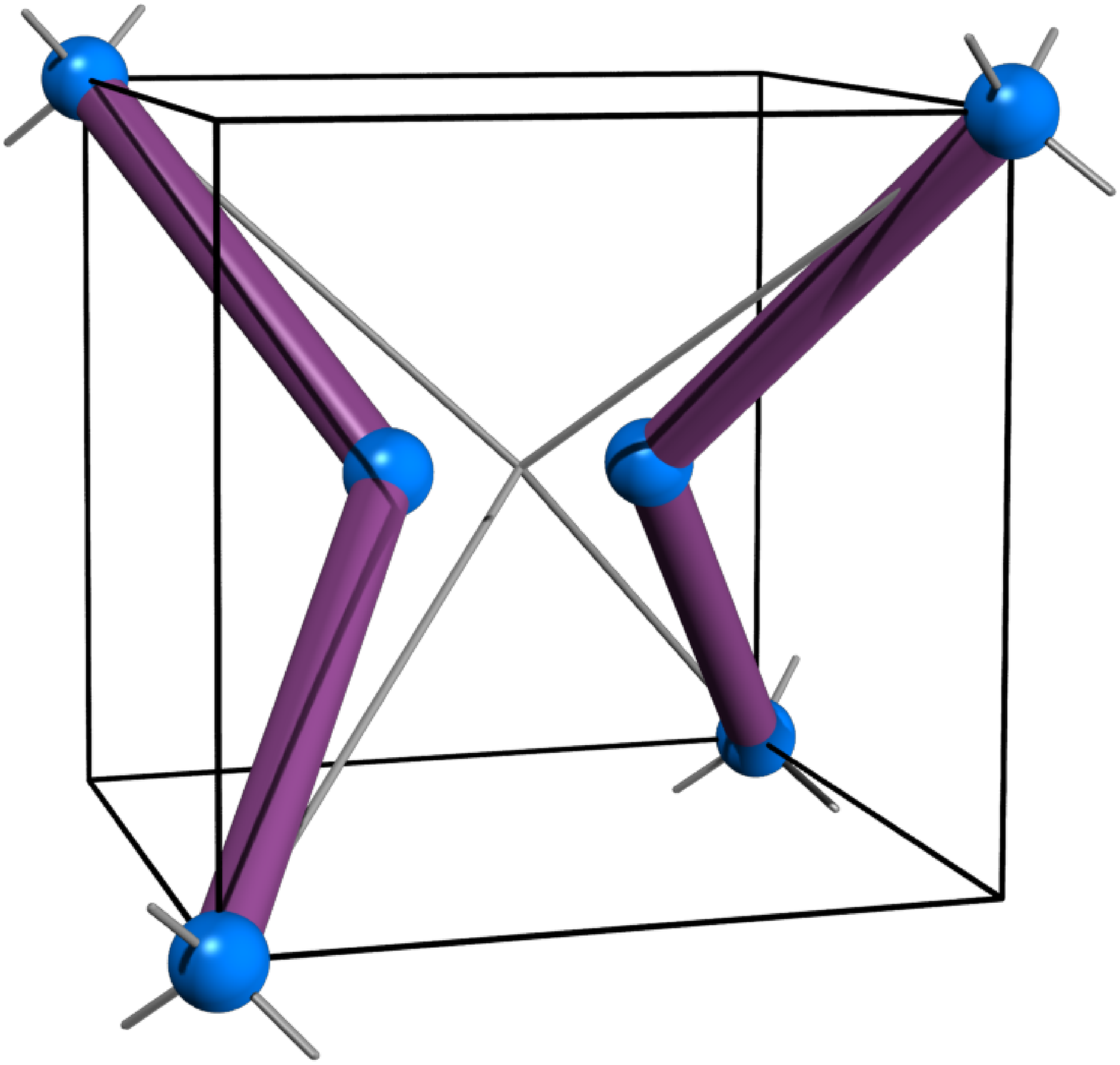}\nolinebreak\hspace*{0.01\columnwidth}
  \includegraphics[width=0.3\columnwidth]{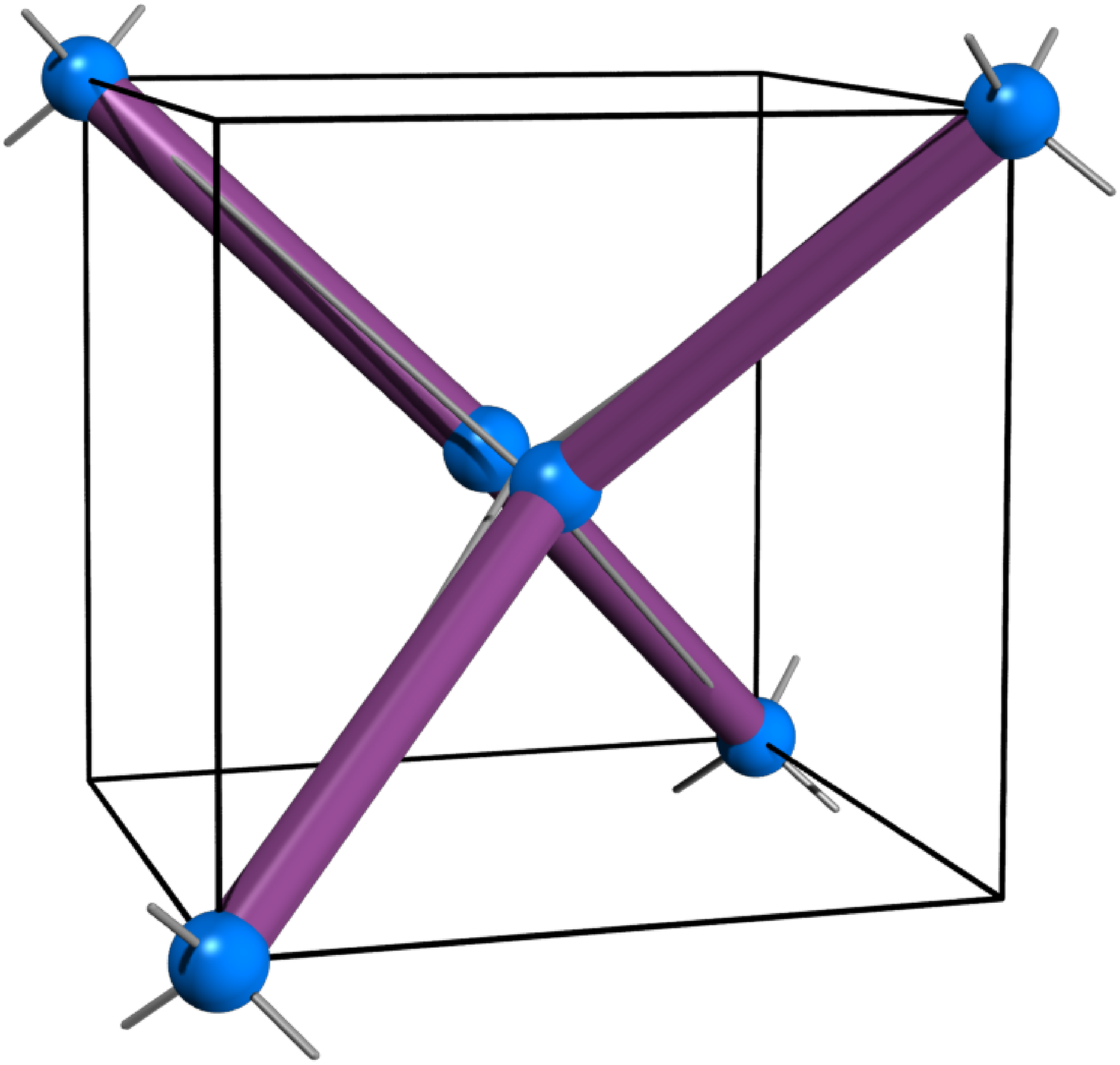}\nolinebreak\hspace*{0.01\columnwidth}
    \caption{A unit cell of the crystallographic diamond network with four-coordinated nodes (top, left) and the network that results from severing one of the nodes (top, right). Each node remains unchanged with probability $(1-p)$, or is severed (or split) into two pairs of edges, with the three possible configurations for neighbour pairs selected with equal probability $p/3$ (bottom).}
\label{fig:NetworkUnlinking}
\end{center}
\end{figure}

We propose a simple statistical model, here referred to as {\em vertex split model} or {\em linked loop model}, defined for the three-dimensional diamond network \footnote{The diamond network is the crystallographic net with  cubic symmetry $Fd\overline{3}m$ consisting of a single type of edge and vertex. 4 edges meet at every vertex, forming tetrahedral angles \cite{DelgadoFriedrichsOKeeffeYaghi:2003c}.}. Rather than deleting spatial elements from the diamond network (such as bonds or vertices), the random operation consists of reducing the vertex coordination by replacing, with probability $p$, each four-coordinated vertices with pairs of two-coordinated vertices, see Fig.~\ref{fig:NetworkUnlinking}. This induces a transition from a fully coordinated crystalline network at $p=0$ to a network filled densely with self-avoiding random walks. The two names are motivated by two different perspectives; with reference to the ordered fully-connected crystalline diamond network at $p=0$, {\em vertex splitting} is the operation that leads to the transition studied here. From the alternative perspective of the state at $p=1$, represented by a dense set of self-avoiding walks, the model may be defined as the random insertion of 'links' between adjacent, infinite or finite loops with probability $(1-p)$.

Each unsevered vertex of the diamond network has four edges connecting the vertex to four distinct neighbour vertices. Each vertex of the diamond network is split (or severed) with probability $p$, that is, the four-coordinated node is replaced by two two-coordinated nodes slightly displaced from the position of the original four-coordinated node, see Fig.~\ref{fig:NetworkUnlinking}. When splitting a node, the three possible configurations for neighbour pairs are selected with equal probability. Note that the parameter $p$ is the probability to {\em degrade} a four-link, opposite to the conventional use of $p$ in bond/site percolation models as the probability to {\em create} a bond or site.

\begin{figure}[t]
\begin{center}
\vspace*{-0.35cm}
\includegraphics[angle=270,width=0.92\columnwidth]{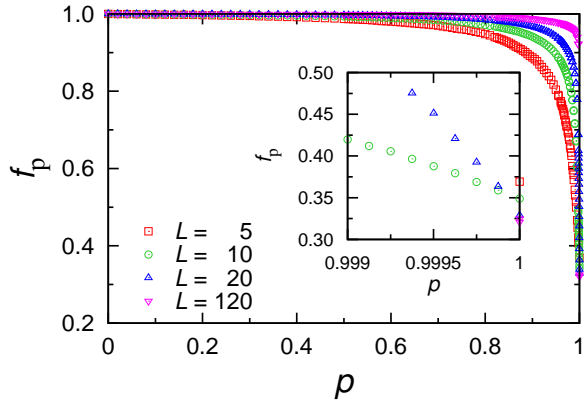}
\caption{Fraction~$f_p$ of edges belonging to any of the percolating clusters. The insert is a close-up, demonstrating a finite value $f_p^c$ of\ $f_p$ at $p_c=1$. All data in Figs.~\ref{fig:fracPercNodes} to \ref{fig:elastic-percolation-exponent} is obtained with lateral periodic boundary conditions.}
\label{fig:fracPercNodes}
\end{center}
\end{figure}

\begin{figure}[t]
  \begin{center}
\vspace*{-0.35cm}
    \includegraphics[angle=270,width=0.92\columnwidth]{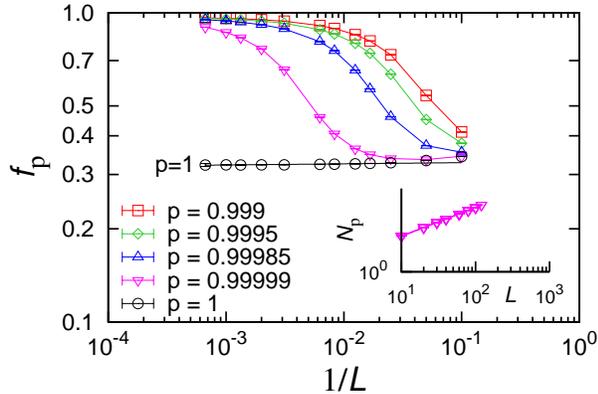}
    \caption{Dependence of~$f_p$ on the linear system size $L$, for fixed $p$. ($8\times L^3$ is the number of vertices.) The insert shows the number $N_p$ of percolating clusters as function of $L$. Note that $\beta\approx 0$ implies a finite value of $f_p^c$, even for large $L$.}
    \label{fig:finiteSizeScaling}
  \end{center}
\end{figure}
\begin{figure}[t]
\vspace*{-0.35cm}
\includegraphics[angle=270,width=0.92\columnwidth]{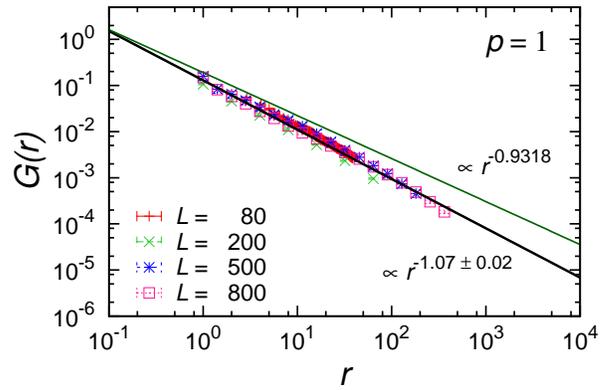}
\caption{The averaged pair-connectedness function $G(r)$, for $p=p_c$, for different linear system size $L$. The decay follows a power law with exponent $(d-2-\eta)=-1.07\pm0.04$, significantly different from the site percolation value $-0.9318$.}
\label{fig:correlationFunction}
\end{figure}

\begin{figure}[t]
\vspace*{-0.35cm}
\includegraphics[angle=270,width=0.92\columnwidth]{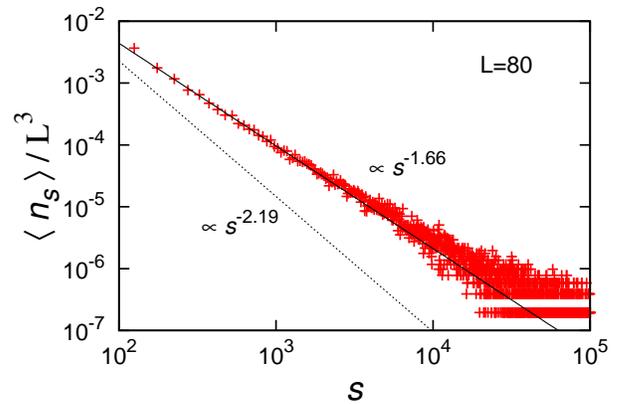}
\caption{Power law decay $\langle n_s\rangle \sim s^{-\tau}$ of the cluster size distribution at $p=p_c$ of the vertex split model with $\tau\approx 1.66$ significantly different from the site percolation exponent $\tau \approx 2.19$ \cite{christensen:2005}. $n_s$ is the number of clusters of size $s$ in a realisation. }
\label{fig:clustersizedistribution}
\end{figure}

\begin{figure}[t]
\vspace*{-0.35cm}
\includegraphics[angle=270,width=0.92\columnwidth]{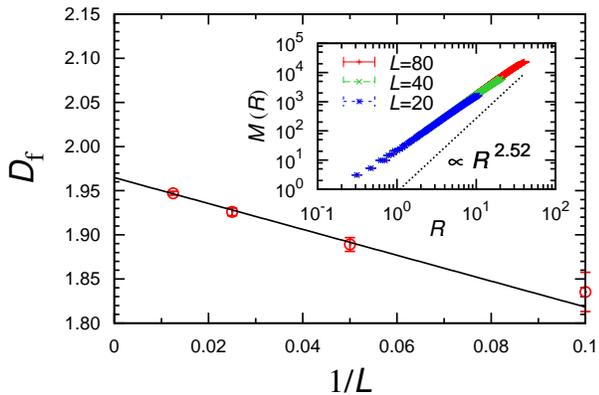}
\caption{The fractal dimension of the percolating cluster at $p=p_c$ is estimated as $D_f=1.96\pm 0.06$. This estimate is obtained by fitting a straight line $a\,R+b$ to the values of average mass of the percolating clusters $M(R)$, see insert for a given system size $L$ \cite{DefinitionMR}. The fractal dimension is obtained by fitting a straight line to the values of $a$ as function of $1/L$. Data in the insert is averaged over 100 or more realisations. The error estimate for $D_f$ corresponds to the variations observed for different system sizes $L=10,20,40,80$ ($L=10$ not shown).}
\label{fig:fractalDimension}
\end{figure}
\begin{figure}
\centering
\vspace*{-0.35cm}
\includegraphics[angle=270,width=0.92\columnwidth]{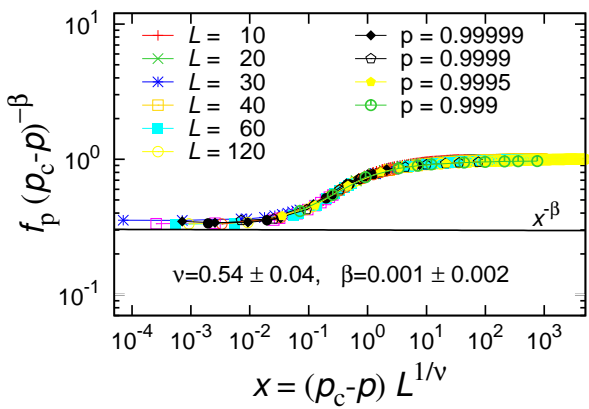}
\caption{Extraction of $\beta$ and $\nu$ from finite size scaling of~$f_p$: the scaling function is expected to become constant for $x=(p-p_c) L^{1/\nu} \gg 1$, and to decrease as $x^{-\beta}$ for $x \ll 1$. Error margins represent the half-width of the intervals for $\beta$ and $\nu$ for which the scaling behaviour is similarly close as for the estimated best values $\nu=0.54$ and $\beta=0.001$. }
\label{fig:fracPercNodesScaling}
\end{figure}

\begin{table}[t]
  \begin{center}
    \begin{tabular}{lr|c|c|c}
      \hline
      & & site & vertex split & Fig.~\\
      \hline
      threshold& $p_c=$ & 0.56988 & 1 & \ref{fig:fracPercNodes} \& \ref{fig:finiteSizeScaling}\\[0.2cm]
      \multicolumn{2}{l|}{edges in infinite cluster(s)} &&&\\
      $f_p\sim (p-p_c)^{\beta} $ & $\beta=$ & $0.41$ & $0.001 \pm 0.002$ & \ref{fig:fracPercNodesScaling} \& \ref{fig:finiteSizeScaling} \\[0.2cm]
      \multicolumn{2}{l|}{correlation length} &&&\\
      $\xi \sim |p-p_c|^{\nu} $& $\nu=$ & $0.88$ &  $0.54\pm 0.04$ & \ref{fig:finiteSizeScaling} \\[0.2cm]
      \multicolumn{2}{l|}{pair-connectedness function} &&&\\
      $G(r) \sim r^{-(d-2+\eta)}$ & $\eta=$ & $-0.068\quad$ & $0.07\pm 0.04$ & \ref{fig:correlationFunction}\\[0.2cm]
      \multicolumn{2}{l|}{cluster size distribution} &&&\\
      $\langle n_s\rangle \sim s^{-\tau}$ & $\tau=$ & $2.18906$ & $1.66\pm 0.14$ & \ref{fig:clustersizedistribution}\\[0.2cm]
      \multicolumn{2}{l|}{fractal dimension} &&&\\
      & $D_{f}=$ & $2.52$ & $1.97\pm 0.06$ & \ref{fig:fractalDimension}\\[0.1cm]
      \hline
    \end{tabular}
    \caption{Thresholds and exponents of the vertex split model and of the site percolation model, both on the diamond network. The critical exponents for the site percolation model are values from refs.~\cite{StaufferAharony:1994,torquato:2002,christensen:2005}, with the percolation threshold for the diamond network reproduced from \cite{vanDerMarck:1998}. For $\beta$ and $\nu$ we have verified that our implementation of the site percolation model reproduces these results. For the scaling exponents of the vertex split model, error bars combine variances of the data (statistical error of fit) with variations when changing fitting ranges and system size. All data is for periodic boundary conditions in the lateral directions; see ref.~\cite{Nachtrab:2011} for data for open boundary conditions.}
    \label{tab:criticalExponents}
  \end{center}
\end{table}

\begin{figure}[t]
\vspace*{-0.35cm}
\includegraphics[angle=270,width=0.92\columnwidth]{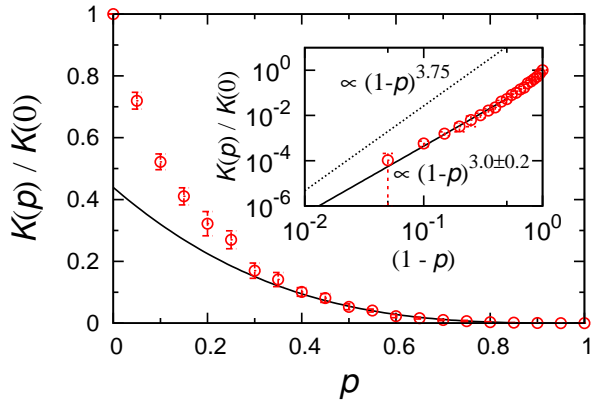}
\caption{Bulk modulus $K(p)$ as a function of $p$ for the network solids obtained by dilating edges to cylinders, with solid volume fraction $\phi=0.1$. The only relevant microscopic linear-elastic material constant is chosen as the Poisson's ratio $\nu_m=0.5$. Data is computed for network solids of $4^3$ unit cells (512 vertices), discretised by $200^3$ voxels and averaged over 5 independent realisations. The insert shows that, near $p_c=1$, the data follows a power-law with exponent $\approx 3.0$ (determined by straight-line fitting to all data for $p\in [0.1,0.5]$), different from the site percolation value $f_c=3.75$ \cite{torquato:2002}}
\label{fig:elastic-percolation-critical-exponent}
\end{figure}

\begin{figure}[t]
\vspace*{-0.35cm}
\includegraphics[angle=270,width=0.92\columnwidth]{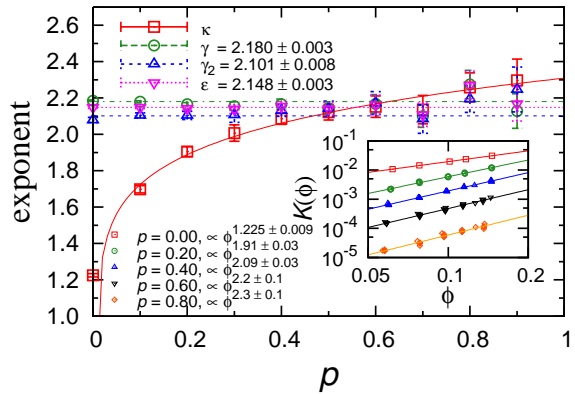}
\caption{For fixed $p$, the effective bulk modulus $K$ and the shear moduli $G_1$ and $G_2$ obey power-laws as function of solid volume fraction $\phi$, see insert. The exponents for the shear moduli, $G_i\propto \phi^{\gamma_i}$ with $i=1,2$ are found to be close to the literature value $2$, and constant as function of $p$. By contrast, the exponent for the bulk modulus $K\propto \phi^{\kappa}$ changes from the expected value $\kappa\approx 1$ at $p=0$ to $\kappa \approx 2$ for $p\approx 1$.}
\label{fig:elastic-percolation-exponent}
\end{figure}

Connected components define clusters, the size of which is measured by the number of constituent edges. A cluster is considered {\em percolating} if it traverses the system in $z$-direction from top to bottom \footnote{Note however that the network itself has cubic symmetry and the severing process induces no anisotropy. In fact, in terms of the effective linear-elastic properties, the system becomes more isotropic with increasing $p$, such that the difference between the two shear moduli (in a system with initially cubic symmetry) vanishes for large $p$, see insert in Fig.~3B in ref.~\cite{Nachtrab:2011b}. This mechanical isotropy relates presumably also to higher structural isotropy, in a statistical sense.}. Systems with both periodic and open boundary conditions in the lateral $x$ and $y$ directions are considered, but all data in Figs.~\ref{fig:fracPercNodes} to \ref{fig:elastic-percolation-exponent} are for periodic boundary conditions. System size is measured in the number $L^3$ of unit cells, each comprising $8$ vertices of the network, see Fig.~\ref{fig:NetworkUnlinking}.

Figure \ref{fig:fracPercNodes} shows the fraction of percolating edges~$f_p$ as function of $p$, for different system sizes $L$, suggesting a transition at $p_c=1$. The entire interval $p\in [0,1)$ represents a percolating phase. At the critical point $p_c=1$ the system still percolates, with the fraction $f_p$ of the system that is part of percolating clusters being a finite constant $f_p^c$; the model has no non-percolating phase. $f_p$ is defined as the ratio $s_p/S$ of the number of all edges that belong to percolating clusters, $s_p:=\sum_{j\in \mathcal{P}} s(j)$, where $\mathcal{P}$ is the set of all percolating clusters and $s(j)$ the number of edges in cluster $j$, to the total number $S$ of edges in the system.

Figure \ref{fig:finiteSizeScaling} shows $f_p$ as function of system size $L$, for various $p$ near $p_c$ and for periodic boundary conditions. This demonstrates that for $p<p_c=1$ the percolating fraction $f_p$ of the system increases with $L$; for large systems and $p<1$ a vast majority of the edges belong to percolating clusters. By contrast, at $p=p_c=1$, $f_p$ varies only very slightly with $L$, despite our analysis including very large systems with up to $2.7\times 10^{10}$ vertices. Linear regression yields a value $\beta/\nu=0.002\pm 0.002$ for the power-law $f_p\propto L^{-\beta/\nu}$, compatible also with $f_p(L)=\mathrm{const}$ \footnote{Error margins combine the statistical variance of the data around their mean and systematic variations observed when varying the fit interval}. This is further support of our claim of a transition at $p_c=1$. The data for open boundary conditions in the lateral directions (shown in \cite{Nachtrab:2011}) are qualitatively similar, yet with the absolute values of $f_p^c$ approximately a factor of 10 smaller. Note in particular that $\beta\approx 0$ implies a finite value of $f_p(p_c)=f_p^c$, even for $L\rightarrow \infty$, in contrast to the standard percolation models.

Figure \ref{fig:finiteSizeScaling} also shows the number of percolating clusters at $p=p_c=1$ as a function of system size $L$. Importantly, in contrast to bond or site percolation, the number of percolating clusters at $p_c$ is not $1$, but grows (within the limits of our numerical resolution) linearly with system size, $N_p\propto L$. The three possible types of unbranched self-avoiding paths for systems with lateral periodic boundary conditions are closed loops, percolating clusters (traversing the system in $z$-direction) and {\em u-turns}, i.e.~clusters that return to the same end (bottom or top) of the network from where they emanated. The probability that a cluster emanating from one of the $4L^2$ sites at $z=0$ percolates appears to be anti-proportional to the system height, $\propto 1/L$. As any u-turn cluster occupies two sites at $z=0$, $N_p$ must be even or zero.

Figures \ref{fig:finiteSizeScaling}, \ref{fig:correlationFunction}, \ref{fig:clustersizedistribution}, \ref{fig:fractalDimension} and \ref{fig:fracPercNodesScaling} support the claim that the transition at $p_c=1$ is a phase transition with scaling behaviour given by power-law decay for the characteristic quantities listed in Tab.~\ref{tab:criticalExponents}. The scaling exponents are significantly different from the critical exponents of conventional bond or site percolation, substantiating the claim that the transition of the vertex split model is different from the universality class of standard percolation.

Several aspects of the model deserve further scrutiny. First, the scaling exponents of the vertex split model do not fulfil the scaling relations $d-D_f=\frac{\beta}{\nu}$ and $d-2+\eta=\frac{2\beta}{\nu}$, valid for bond or site percolation; their derivation assumes a single unique percolating cluster, in contrast to the many line-like percolating clusters in this model. 

Second, the corresponding planar model of severing four-coordinated vertices of planar square lattices is closely related to hull percolation or hull exponents of standard percolation clusters \cite{Ziff:1986,*DuplantierSaleur:1987,*RouxGuyonSornette:1988,*AizenmanDuplantierAharony:1999}, also with different critical behaviour from standard percolation; the relationship between the planar and the spatial case requires further exploration. 

Third, the values of some of the scaling exponents, in particular $\beta\approx 0$, point to the possibility of an effective description of the system. The perspective of the 'loop link' model affords the interpretation that, at $p=1$, the insertion of random links between adjacent pairs of self-avoiding random walks corresponds to a long-range effect, which induces a sharp (possibly first order) transition with $\beta=0$. This, as well as the link to Flory-type arguments for the scaling behaviour of polymer systems, requires further investigation, by models that effectively tune the characteristics of the self-consistent random walk configuration.

Finally, in analogy to standard percolation, one may expect the critical behaviour to be independent of the type of underlying network; this expectation could be verified by an analysis of node severing of other four-coordinated networks, such as the crystalline {\bf nbo} network \cite{DelgadoFriedrichsOKeeffeYaghi:2003c} or the network of Plateau edges in random foams \cite{KraynikReineltVanSwol:2004}. 


The remainder of this paper addresses mechanical properties of linear-elastic realisations of the networks with split (or severed) vertices. In porous or cellular structures, the existence of a solid percolating cluster is a prerequisite for mechanical stability, that is, for finite values of the effective linear-elastic moduli. The relationship between percolation critical behaviour and effective elastic properties (those relevant for sample sizes much larger than the microstructural length scale) is well-known, leading to a power-law decay of the effective elastic moduli near $p_c$ \cite{sahimi:1994,torquato:2002}. We employ a voxel-based finite element method \cite{Nachtrab:2011a,*KapferBiomaterials:2011} to evaluate the effective linear-elastic properties of network solids based on the vertex split model \footnote{Some preliminary results, for $p\ll p_c$ far from the percolation critical point, have been published in~\cite{Nachtrab:2011b}}.

Figure \ref{fig:elastic-percolation-critical-exponent} shows that near $p_c$, the effective bulk modulus $K$ (the resistance to hydrostatic compression) is commensurate with a power-law decay, $K \propto |p-p_c|^{f_c}$, with an exponent $f_c\approx 3.0$, significantly different from the known exponent $f_c=3.75$ \cite{torquato:2002} for site percolation \footnote{An analysis of site percolation with the FEM scheme used yields $f_c=3.6\pm 0.1$.}. 

The change in network structure that occurs as $p$ varies from $0$ to $1$ is reflected in the density dependence of the linear elastic bulk modulus. It is frequently observed that the effective elastic moduli scale as power-laws in the solid volume fraction $\phi$ of the cellular structure; specifically, for the limit $\phi\rightarrow 0$ of thin beams, the bulk modulus follows $K\propto \phi^\kappa$ with $\kappa=1$ and the shear moduli \footnote{Note that structures with cubic symmetry, such as the crystallographic diamond network, have three independent elastic moduli, the bulk modulus $K$ and two shear moduli $G_1$ and $G_2$.} $G\propto \phi^\gamma$ with $\gamma=2$ \cite{GibsonAshby:1982}. For the vertex split model, Fig.~\ref{fig:elastic-percolation-exponent} shows that the effective exponent $\kappa$ of the bulk modulus varies from a value near $1$ (as expected) at $p=0$ to a value close to $2$ when all nodes are disconnected at $p=1$. The exponents of the shear moduli remain close to the expected value of $2$.

This behaviour is somewhat rationalised by the observation that, in ordered cellular structures in the thin beam limit, linear behaviour of elastic moduli is associated with strut compression being the dominant deformation mode, whereas quadratic behaviour is associated with strut bending or torsion \cite{WarrenKraynik:1988,*WarrenKraynik:1997,*Christensen:2000}. The network solids corresponding to the vertex split model appear to undergo a transition from being compression-dominated when fully four-coordinated at $p=0$ to being bending-dominated in the terminal state (at $p=1$) which corresponds to a dense set of self-avoiding polymers.  

In conclusion, we have demonstrated that randomly severing the four-coordinated vertices of a diamond network leads to a transition, manifest in the fraction of clusters that are percolating. The transition, which is reminiscent of a percolation transition yet with substantially different behaviour to conventional bond/site percolation, occurs at $p_c=1$ when {\em all} nodes have been split.

\begin{acknowledgments}
We thank Andrew M.~Kraynik for advise on mechanical properties of cellular structures, and G.\ Last and S.\ Ziesche for insightful discussion of percolation transitions. We acknowledge the support of the German Science Foundation (DFG) through the Cluster of Excellence 'Engineering of Advanced Materials' and through the research group ``Physics and Geometry of Random Spatial Structure'' under grant ME1361/12-1.
\end{acknowledgments}

\end{document}